\documentclass[twocolumn,10pt,A4paper]{IEEEtran}

\usepackage{comment}
\includecomment{toexclude}
\usepackage{amsmath,amsfonts,amssymb}
\usepackage{cite}
\usepackage{verbatim}
\usepackage{bm}
\usepackage{algorithm}
\usepackage{graphicx}
\usepackage{subfigure}
\usepackage{ragged2e}
\usepackage{tabularx}
\usepackage{array}
\usepackage{multirow,multicol} 
\usepackage{url}

\usepackage{color}
\usepackage{xcolor}
\graphicspath{{./image/}}

\newcommand{\cb}{} 

\newcommand{\wcf}{\mathbf{w}_{t,c}}
\newcommand{\wsf}{\mathbf{w}_{t,s}}
\newcommand{\wtf}{\mathbf{w}_t}
\newcommand{\wrf}{\mathbf{w}_r}
\newcommand{\wf}{\mathbf{w}}

\newcommand{\qf}{\mathbf{q}}
\newcommand{\wtft}{\tilde{\mathbf{w}}_t}
\newcommand{\af}{\mathbf{a}}

\newcommand{\pf}{\mathbf{p}}
\newcommand{\df}{\mathbf{d}}

\newcommand{\Lt}{\tilde{L}}

\newcommand{\etat}{\tilde{\eta}}

\newcommand{\Ht}{\mathbf{H}}
\newcommand{\At}{\mathbf{A}}
\newcommand{\Dt}{\mathbf{D}}

\newcommand{\It}{\mathbf{I}}

\newcommand{\Wt}{\mathbf{W}}

\DeclareMathOperator*{\argmin}{arg\,min}
\DeclareMathOperator*{\argmax}{arg\,max}

\usepackage[outdir=./]{epstopdf}
\graphicspath{{./image/}}

\DeclareMathOperator{\Tr}{Tr}

\def\BibTeX{{\rm B\kern-.05em{\sc i\kern-.025em b}\kern-.08em
		T\kern-.1667em\lower.7ex\hbox{E}\kern-.125emX}}
\begin{document}
	
\title{Multibeam Optimization for Joint Communication and Radio Sensing Using Analog Antenna Arrays}

\author{Yuyue~Luo,~\IEEEmembership{Student Member,~IEEE}, J.~Andrew~Zhang,~\IEEEmembership{Senior Member,~IEEE},\\ Xiaojing~Huang,~\IEEEmembership{Senior Member,~IEEE}, Wei~Ni,~\IEEEmembership{Senior Member,~IEEE}, and Jin~Pan
	\thanks{This work was supported by the Foundation for Innovative Research Group of the National Natural Science Foundation of China under Grant No. 61721001. \textit{(Corresponding author: Jin Pan)}}
	\thanks{Yuyue Luo and Jin Pan are with School of Electronic Science and Engineering, University of Electronic Science and Technology of China, China. Yuyue Luo is also with School of Electrical and Data Engineering, University of Technology Sydney, Australia. Email: Yuyue.Luo@student.uts.edu.au; jpuestc@163.com.}
	\thanks{J. Andrew Zhang and Xiaojing Huang are with School of Electrical and Data Engineering, University of Technology Sydney, Australia. Email:\{Andrew.Zhang;Xiaojing.Huang\}@uts.edu.au.}
	\thanks{Wei Ni is with Data61, CSIRO, Sydney, Australia, NSW 2122. E-mail: wei.ni@data61.csiro.au.}
	\thanks{Part of the work presented in this paper was accepted for publication in IEEE GLOBECOM2019. More than $50\%$ of the work here is new and different to that paper.}

}
\markboth{Journal of \LaTeX\ Class Files,~Vol.~14, No.~8, August~2015}%
{Shell \MakeLowercase{\textit{et al.}}: Bare Demo of IEEEtran.cls for IEEE Journals}

\maketitle

\begin{abstract}
Multibeam technology enables the use of two or more subbeams for joint communication and radio sensing, to meet different requirements of beamwidth and pointing directions. Generating and optimizing multibeam subject to the requirements is critical and challenging, particularly for systems using analog arrays. This paper develops optimal solutions to a range of multibeam design problems, where both communication and sensing are considered. We first study the optimal combination of two pre-generated subbeams, and their beamforming vectors, using a combining phase coefficient. Closed-form optimal solutions are derived to the constrained optimization problems, where the received signal powers for communication and the beamforming waveforms are alternatively used as the objective and constraint functions. We also develop global optimization methods which directly find optimal solutions for a single beamforming vector. By converting the original intractable complex NP-hard global optimization problems to real quadratically constrained quadratic programs, near-optimal solutions are obtained using semidefinite relaxation techniques. Extensive simulations validate the effectiveness of the proposed constrained multibeam generation and optimization methods.
\end{abstract}


\section{Introduction} \label{sec-intro}
Joint communication and radio sensing (JCAS) techniques, also known as Radar-Communications, have received increasing interest from both academia and industry \cite{sturm2011waveform,han2013joint,hassanien2016signaling,petrov2019unified,kumari2018ieee}. It has appealing features, such as low cost, resource saving, reduced size and weight, and mutual sharing of information, for improved communication and sensing performance \cite{zhang2019multibeam,luo2019optimization}. Millimeter-wave (mmWave) JCAS systems can potentially provide very high data-rate communications and high accurate sensing results, due to their large signal bandwidth and small-profile massive antenna arrays.

Steerable beamforming (BF) technique can overcome large propagation attenuation, supporting mobility and exploiting channel sparsity in mmWave JCAS. However, there are challenges associated with the technique, particularly in systems using a single analog array. The primary challenge is that communication and sensing have different requirements for BF. Radio sensing often requires time-varying directional scanning beams, while a stable and accurately-pointing beam is usually expected for communication. In \cite{kumari2018ieee,dokhanchi2019mmwave,blunt2010embedding,hassanien2015dual,hassanien2016non}, a single beam was used for communication and sensing, and hence sensing is restricted within the communication direction.

\cb{Multibeam technology \cite{hong2017multibeam} which enables the use of BF waveform with more than one mainlobe (called as subbeam hereafter), has a wide range of applications, such as radar \cite{steyskal1989digital,pfeffer2013fmcw}, satellite communications \cite{vazquez2016precoding,atesal2011two}, wireless communications \cite{wei2018multi} and radio astronomy \cite{warnick2016high}. Recently, multibeam technology, as a viable solution to the BF problem, has been applied to JCAS, such as the use of BF network circuit \cite{nusenu2019dual,mccormick2017simultaneous} and digital BF in MIMO systems \cite{wang2018spatial,liu2018mu,liu2018toward}.} In \cite{wang2018spatial}, sparse antenna array and BF optimization were studied for communication-embedded MIMO radar systems. In \cite{liu2018mu}, multibeam waveform optimization was designed to minimize the difference between the generated and the desired sensing waveforms under the constraints on the signal-to-interference-and-noise ratio (SINR) of multiuser MIMO communications. In \cite{liu2018toward}, globally optimal waveforms were derived for multiple desired radar beam patterns, based on the criterion of minimizing multiuser interference for communications. These solutions were based on digital MIMO systems, which are not always feasible for mmWave due to high hardware complexity and cost.
More cost-effective options for mmWave JCAS were suggested to be analog or hybrid arrays \cite{heath2016overview}. 

A multibeam scheme for JCAS with analog arrays was first introduced in \cite{zhang2017framework} and then improved in \cite{zhang2019multibeam,luo2019optimization}. In that scheme, the multibeam consists of a fixed subbeam dedicated to communication along and a scanning subbeam with a direction varying across different packets. Several methods for generating the multibeam varying over packets were proposed in \cite{zhang2019multibeam}. Method 2 in \cite{zhang2019multibeam} directly generates the multibeam by minimizing the mismatches between the desired and the generated BF waveforms using an iterative least squares (ILS) method, without consideration on communications. Method 1 in \cite{zhang2019multibeam} is a low-complexity and flexible \textit{subbeam-combination} method, where two basic beams for communication and sensing are separately generated according to the desired BF waveform. The two beams are further shifted to the desired directions by multiplying a sequence, and then combined by using a power distribution factor and a phase shifting coefficient. Method 1 also provides a simple way to determine the phase shifting coefficient, ensuring that the fixed and scanning subbeams have the same phase in the dominating communication direction and can be combined constructively for communication, with no consideration on the sensing waveform. Closed-form optimal solutions for the phase coefficient, as well as for the quantization of the BF vectors, were further investigated in \cite{luo2019optimization}. \cb{The analog rank-one multibeam BF with a single RF chain was proved to be simple, cost-effective, compact, and computationally efficient \cite{zhang2019multibeam,luo2019optimization}, and is suitable for portable applications such as JCAS in unmanned aerial and ground vehicles.}

Two remaining important issues are yet to be addressed in multibeam design \cite{zhang2019multibeam,luo2019optimization}. Firstly, in \cite{luo2019optimization}, the optimization of the combining coefficient was conducted by maximizing the received signal power at the communication receiver, without explicit consideration on the sensing waveform. Although the impact was demonstrated to be statistically small via numerical simulations, the waveform at the sensing directions can distort occasionally. Secondly, although the subbeam-combination method investigated in \cite{zhang2019multibeam, luo2019optimization} is simple and flexible for implementation, it is suboptimal because the BF weights are separately pre-generated for the two subbeams and combined by only a single variable. It is unclear what its performance gap is from the optimum and whether the latter exists. 

In this paper, we propose new multibeam optimization techniques which take into account both communication and sensing performance of a JCAS system with analog arrays, hence addressing both of the above issues comprehensively. We are particularly interested in two classes of optimization problems: 1) maximizing the received signal power for communications subject to the constraints on the scanning subbeam; and 2) optimizing the BF waveform with constraints on the received signal power for communications. For both problems, we first study the subbeam combination method in \cite{zhang2019multibeam, luo2019optimization} but with new holistic analysis and solutions developed, and then design the global optimization techniques. Our main contributions are summarized as below.
\begin{itemize}
\item For the subbeam-combiner method, we propose new approaches to maximize the received signal power in the cases of (1) constrained BF gain at discrete scanning directions and (2) constrained total scanning power over a range of directions. In both cases, we show that closed-form optimal solutions for the combining coefficient can be obtained by finding common solutions to the objective and constraint functions;
\item For the subbeam-combiner method, we provide closed-form constrained optimal solutions that maximize the scanning gain in particular directions or the scanning power over a given range of directions, subject to the constraint on the received signal power for communications. This is a dual problem to that stated in the contribution above. These optimal solutions, as well as those above, are shown to be practical and efficient, and can be obtained at low computational complexities;  
\item We develop new global optimization methods that directly optimize the BF vector, considering the requirements of both communication and sensing. We introduce a novel method to convert the original NP-hard complex problems to real quadratically constrained quadratic programs (QCQPs), which are then solved efficiently by semidefinite relaxation (SDR) techniques \cite{ma2010semidefinite}. These methods achieve near-optimal solutions, providing benchmarks for performance evaluation of suboptimal solutions.
\end{itemize}
Extensive simulation results validate the effectiveness of the proposed BF optimization methods.

The rest of this paper is organized as follows. We introduce the system model, formulate the problems, and elaborate on our principle of multibeam optimization in Section \ref{sec-Foundation}. Constrained optimization methods for the combining coefficients are investigated in Section \ref{sec-OptPhi}. The proposed global optimization methods are described in Section \ref{sec-SDR}. In Section \ref{sec-simulation}, extensive simulation results are presented, and finally, concluding remarks are provided in Section \ref{sec-conc}.

Notations: $(\cdot)^H$, $(\cdot)^*$, $(\cdot)^T$, $(\cdot)^{-1}$, and $(\cdot)^\dagger$ denote Hermitian transpose, conjugate, transpose, inverse, and pseudo-inverse, respectively. $|\cdot|$ and $\|\cdot\|$ denote element-wise absolute value and Euclidean norm, respectively. $E(\cdot)$ denotes expectation. $\arg(\cdot)$ denotes the argument of a complex number. $\mathbb{R}^n$ and $\mathbb{S}^n$ denote the sets of all real $n\times n$ matrices and real symmetric $n\times n$ matrices, respectively.

\section{System Model}\label{sec-Foundation}

In this paper, we consider the same system set-up as in \cite{zhang2019multibeam}. Two nodes perform two-way point-to-point communications in time division duplex (TDD) mode and simultaneously sensing the environment to determine the locations and speed of nearby objects. To mitigate the leakage from the transmitter to the receiver, each node uses two spatially separate analog antenna arrays, for transmission and reception, respectively. \cb{Each analog array only has a single radio frontend (RF) chain. The received signals at the different antennas of the receiver array are weighted, and combined before being sent to an analog-to-digital converter (ADC). The digital baseband signal is converted to analog by a digital-to-analog convertor (DAC), and then weighted and fed to the different antennas of the transmitter array.  The weighting, with a complex value, can be achieved using analog circuits either passively \cite{luo2019optimization} or actively \cite{naqvi2018review}.} Below we briefly describe the system. The readers are referred to \cite{zhang2019multibeam} for more details of the system and multibeam JCAS technology. 

We consider $M$-element uniform linear arrays (ULAs) with half-wavelength antenna spacing. Considering planar wave-front and a narrow-band BF model, the array response vector is given by
\begin{align}\label{eq-arrayModel}
\af(\theta)=[1,e^{j\pi \sin(\theta)},\cdots, e^{j\pi (M-1)\sin(\theta)}]^T, 
\end{align}
where $\theta$ is either the angle-of-arrival (AoA) or angle-of-departure (AoD).

\cb{Similar to \cite{heath2016overview,alkhateeb2014channel,el2014spatially,alkhateeb2015limited}, this work considers a narrowband beamforming model and a narrowband sparse channel model with a dominant line-of-sight (LOS) path and a limited number of much weaker non-line-of-sight (NLOS) paths. On one hand, the validity of the narrowband beamforming model relies on the fractional bandwidth, which is defined as the ratio between signal bandwidth $W$ and carrier frequency $f_c$. When the fractional bandwidth is sufficiently small, i.e., $W/f_c\ll1$, the variation of the phase shift across different frequencies, i.e., the beam squint effect, is ignorable and the narrowband beamforming model is valid \cite{heath2016overview}. On the other hand, in a typical mmWave environment, the power ratio between the LOS and NLOS paths is typically very large \cite{rappaport2013millimeter,maccartney201473}. For example, referring to the measurement channel data for a typical urban environment when the carrier frequency is 73 GHz \cite{maccartney201473}, the power ratio between LOS and NLOS paths is more than 30 dB when the Tx-Rx separation distance is 100 m. 
Therefore, frequency selectivity is negligible and the consideration of a narrowband channel model is reasonable in this paper. In particular, all multipath signals are assumed to cause negligible inter-symbol interference in communications. Consider an $L$-path channel with AoDs $\theta_{t,\ell}$ and AoAs $\theta_{r,\ell}$, $l=1,\cdots,L$. The quasi-static physical channels \cite{heath2016overview} can be represented as
	\begin{align}
	\Ht=\sum_{\ell=1}^L b_\ell \delta(t-\tau_{\ell})e^{j2\pi f_{D,\ell} t} \af(\theta_{r,\ell})\af^T(\theta_{t,\ell}),
	\label{eq-Ht}
	\end{align}
	where, for the $\ell$-th path, $b_{\ell}$ is its amplitude, $\tau_{\ell}$ is the propagation delay, and $f_{D,\ell}$ is the associated Doppler frequency.}

Let $s(t)$ be the transmitted baseband signal, and $\wtf$ and $\wrf$ be the transmitter and receiver BF vectors, respectively. The received signal for either sensing or communication can be written as:
\cb{
\begin{align}
\begin{split}
&y(t)=\wrf^T\Ht \wtf\,s(t) + \wrf^T\bm{z}(t)\\
&=\sum_{\ell=1}^L b_\ell e^{j2\pi f_{D,\ell} t} \big(\wrf^T\af(\theta_{r,\ell})\big)\big(\af^T(\theta_{t,\ell})\wtf\big)  s(t-\tau_{\ell})\\
&\ \ \ \ + \wrf^T\bm{z}(t),
\end{split}
\end{align}}where $\bm{z}(t)$ is the additive white Gaussian noise (AWGN) vector at the receiver.

We assume that $\Ht$ is known at the transmitter, and design the BF weight vector $\wtf$ that generates a fixed subbeam in the principal communication direction and a scanning beam for sensing in different directions. The scanning subbeams are designed to scan areas in different directions from the principal communication direction. Both subbeams contain the same information, and are used for both communication and sensing. Hereafter, we call these two subbeams as \textit{fixed} and \textit{scanning} subbeams. For $\wrf$, we assume that maximal ratio combining (MRC) \cite{lo1999maximum} is applied in the analog domain, to achieve the maximal output power at the receiver (or in other words, the maximal SNR). Therefore, $\wf_r=(\Ht\wf_t)^*$.

In \cite{zhang2019multibeam,luo2019optimization}, two BF vectors, $\wcf$ and $\wsf$, are designed to generate the fixed and scanning subbeams, respectively. They are combined by a phase shifting coefficient $e^{j\varphi}$ and a power distribution factor $\rho$ $(0<\rho<1)$, as given by
\begin{align}\label{eq-wt}
\mathbf{w}_t=\sqrt{\rho}\wcf+\sqrt{1-\rho} e^{j\varphi}\wsf.
\end{align}
The value of $\rho$ can typically be determined by balancing the communication and sensing distances \cite{luo2019optimization}. \cb{The optimization is conducted with respect to $\varphi$, which has a non-negligible impact on the BF performance. This is because when we design the BF vectors of the subbeams, $\wcf$ and $\wsf$ are only respectively optimized for magnitudes of the desired BF waveform with no consideration on phases. The BF gain of the combined multibeam for communication and sensing depends on how these two BF vectors are combined. The optimized $\varphi$ can ensure that the two pre-generated subbeams are coherently combined to form the multibeam. When $\wcf$ and $\wsf$ change, for example, $\wsf$ changes every packet due to the varying AoD of the scanning subbeam, $\varphi$ needs to be accordingly updated.} As mentioned in Section \ref{sec-intro}, several suboptimal methods have been proposed in \cite{zhang2019multibeam, luo2019optimization} to optimize $\varphi$, without explicit consideration on the sensing BF waveform. In the rest of this paper, we present constrained optimization methods for $\varphi$ and globally optimal solution for $\wrf$, given the requirements of both communication and sensing. 

\cb{In the following sections, the problem formulations of BF for JCAS will be proposed. The notations used in the formulations are summarized in Table \ref{tb-variables}.}

\begin{table*}
	\cb{
		\caption{A summary of important notations used in this paper.}
		\centering
		\resizebox{2\columnwidth}{!}{%
			{\begin{tabular}{|c|c|c|l|}
					\hline
					\multirow{6}{*}{$\wf$} & \multirow{6}{*}{BF vectors} & $\wtf$ & TX BF vector\\
					\cline{3-4}
					& &  $\wrf$ & RX BF vector\\				
					\cline{3-4}	
					& &  $\wcf$ & TX BF vector for communications\\
					\cline{3-4}	
					& &  $\wsf$ & TX BF vector for radio sensing\\	
					\cline{3-4}	
					& &  $\wf_t^{(q)}, q=1,2,\cdots,8$ & Optimal BF vectors for the $q$th problem formulation\\	
					\hline	
					\multirow{2}{*}{$\varphi$} & \multirow{2}{*}{Phase shifting coefficient} & $\varphi_\text{opt}$ & Optimal phase shifting value without consideration of constraints \\
					\cline{3-4}
					& &  $\varphi_\text{opt}^{(q)}, q=1,2,\cdots,8$ & Optimal phase shifting value for the $q$th problem formulation\\
					\hline
					\multirow{4}{*}{$\Bbbk$} & \multirow{4}{4cm}{\centering Range of $\varphi$ satisfying constraints} & $\Bbbk_i, i=1, 2, …, N_s$ & Range of $\varphi$ satisfying the $i$th constraint in (5b) \\
					\cline{3-4}
					& & $\Bbbk_s$ & Range of $\varphi$ satisfying (5b)\\
					\cline{3-4}
					& & $\Bbbk_p$ & Range of $\varphi$ satisfying (13b) \\
					\cline{3-4}
					& &$\Bbbk_g$ & Range of $\varphi$ satisfying (17c) \\
					\hline
					\multirow{4}{*}{$\theta$} & \multirow{4}{*}{AoDs/AoAs} & $\theta_{t,l}$ & AoD at the $l$th path, $l=1,2,\cdots,L$ \\
					\cline{3-4}
					& &$\theta_{r,l}$ & AoA at the $l$th path, $l=1,2,\cdots,L$ \\
					\cline{3-4}
					& &$\theta_{s_i}$ & The $i$th sensing AoD with a constraint on the minimum BF gain\\
					\cline{3-4}
					& &$\theta_{s_l},\theta_{s_r}$ & Bounds of the range of AoDs with the constraint of the total power\\
					\hline		
					\multirow{3}{*}{$C$} & \multirow{3}{4cm}{\centering Scaling coefficient for the bounds of the constraints} & $C_{si}$ & The $i$th scaling coefficient to the maximum achievable BF gain  \\
					\cline{3-4}
					& &$C_{sp}$ &The scaling coefficient to the total power over a range of consecutive scanning directions\\
					\cline{3-4}
					& &$C_{p}$ &The scaling coefficient to the received signal power\\
					\hline
					\multirow{6}{*}{$\varepsilon$}& \multirow{6}{4cm}{\centering Bounds of the constraints considering global optimizations} & \multirow{2}{*} {$\varepsilon_w$} \multirow{2}{*}& \multirow{2}{10cm} {The bound of the constraint for mismatches between the generated and the desired BF waveforms}\\
					& & & \\
					\cline{3-4}
					& & \multirow{2}{*} {$\varepsilon_{si}$}& \multirow{2}{10cm}{The $i$th bound of the constraint for BF gain of the subbeam in the direction of interest}\\
					& & & \\
					\cline{3-4}
					& & \multirow{2}{*} {$\varepsilon_{p}$}& \multirow{2}{10cm}{The bound of the constraint for the power over a range of consecutive scanning directions}\\
					& & & \\
					\hline
					
				\end{tabular}} 
				\label{tb-variables}}	}	
	\end{table*}

\section{Constrained Optimal Solutions for $\varphi$} \label{sec-OptPhi}

In this section, we investigate several constrained optimization methods to the design of the BF vector in \eqref{eq-wt}. We consider two types of optimization problems: (1) Maximizing the received signal power for communications subject to BF waveform constraints on scanning subbeams; and (2) optimizing the BF waveform of the scanning subbeam subject to constraints on the received signal power for communications. 

\subsection{Maximizing Received Signal Power with Constraints on Scanning Waveform}
We intend to maximize the received signal power and equivalently the received signal-to-noise ratio (SNR) \cite{luo2019optimization} for communications, while meeting constraints on the BF waveform. We study two types of constraints on the sensing subbeam in the following.

\subsubsection{Constrained BF Gain in Discrete Scanning Directions}\label{sec-ScanGain}

We consider the cases where there are constraints on the minimum BF gain in several sensing directions. 
\cb{Let the threshold in the $i$-th sensing direction $\theta_{s_i}$ be $C^2_{s_i} (1-\rho)M$, where $C_{s_i}\in[0,1]$ is a scaling coefficient, representing the ratio between the gain of the scanning subbeam in the direction of interest and the maximum gain that the array can achieve for sensing, i.e., $(1-\rho)M$. In a practical system, the value of $C_{s,i}$ depends on the specific requirement of the BF gain in the directions of interest, which depends on the radar sensing parameters, such as the desired range of detection and the distance of targets.} We can formulate the constrained optimization problem as 
\cb{
\begin{subequations}
	\begin{align}
	&\text{P}_1:\ {\varphi}^{(1)}_\text{opt}=\arg\max_{\varphi}\dfrac{\wtf^H\Ht^H\Ht\wtf}{\|\wtf\|^2}, \label{eq-ConsA}\\
	&\text{s.t. }\ \dfrac{|\af^{T}(\theta_{s_i})\wf_t|^2}{||\wf_t||^2}\geq C^2_{s_i} (1-\rho)M, \  i=1,2, \cdots, N_s, \label{eq-stscan}\\
	&\text{with}\ \mathbf{w}_t=\sqrt{\rho}\wcf+\sqrt{1-\rho} e^{j\varphi}\wsf, \notag	
	\end{align}
\end{subequations}
where $N_s$ is the number of constraints; and $\wcf$ and $\wsf$ are pre-designed to generate the communication and scanning subbeams with the required BF waveform, respectively.}

Let ${\varphi}_{\text{opt}}$ be the unconstrained optimal solution for \eqref{eq-ConsA}, which was already obtained in \cite{luo2019optimization}. To solve the constrained optimization problem, we can first evaluate the range of ${\varphi}$ for each constraint in \eqref{eq-stscan}, and then check ${\varphi}_{\text{opt}}$ against their intersection. Expanding the left-hand side of the $i$-th inequality of \eqref{eq-stscan}, we obtain
\begin{align}\label{eq-ineq}
&|\af^{T}(\theta_{s_i})\wf_t|^2=\rho|\wcf^H\af^*(\theta_{s_i})|^2+(1-\rho)|\wsf^H\af^*(\theta_{s_i})|^2 \notag\\
&\quad\qquad\qquad\qquad +2P\mathfrak{Re}\{e^{j\varphi}\wcf^H\mathbf{a}^*(\theta_{s_i})\af^T(\theta_{s_i})\wsf\}, \notag\\
&||\wf_t||^2=\rho\|\wcf\|^2+(1-\rho)\|\wsf\|^2+2P\mathfrak{Re}\{e^{j\varphi}\wcf^H\wsf\} \notag\\
&\qquad\quad=1+2P\mathfrak{Re}\{e^{j\varphi}\wcf^H\wsf\},
\end{align}
where $P\triangleq\sqrt{\rho(1-\rho)}$. Let $\wcf^H\wsf=b_1e^{j\beta_1}$, $\wcf^H\af^*(\theta_{s_i})=b_{2i}e^{j\beta_{2i}}$, and $\af^T(\theta_{s_i})\wsf=b_{3i}e^{j\beta_{3i}}$, where the cross-product terms are represented by their magnitude and phase. Further let $B_{1i}\triangleq[\rho b_{2i}^2+(1-\rho)b_{3i}^2]/(2P)$, and $B_{2i}\triangleq MC_{s_i}^2(1-\rho)/(2P)$. Thus each inequality in \eqref{eq-stscan} can be converted to 
\begin{align}
\left\{\begin{array}{l}
X_{1i}\sin\varphi+X_{2i}\cos\varphi\geq {B_{2i}-B_{1i}},\\
X_{1i}\triangleq 2Pb_1B_{2i}\sin\beta_1-b_{2i}b_{3i}\sin(\beta_{2i}+\beta_{3i}),\\
X_{2i}\triangleq b_{2i}b_{3i}\cos(\beta_{2i}+\beta_{3i})-2Pb_1B_{2i}\cos\beta_1.
\end{array}
\right.
\label{eq-convt}
\end{align}
We can now obtain the range of $\varphi$ by considering the following three cases.
\begin{itemize}
\item Case 1): If $|B_{2i}-B_{1i}|\leq\sqrt{X_{1i}^2+X_{2i}^2}$, we can get the solution to \eqref{eq-convt} as a set $\varphi\in\Bbbk_i=\left[\varphi_{1i},\varphi_{2i}\right]$, where $\varphi_{1i}$ and $\varphi_{2i}$ denote the two bounds of the set. The set is given by
\begin{align}\label{eq-ki}
\begin{split}
\Bbbk_i=&\left[\varphi_{1i},\varphi_{2i}\right]\\
=&\left\{\begin{array}{cl}
\left[\mu_i-\sigma_i,-\mu_i+\pi-\sigma_i\right], &\text{if}\ X_{1i}\geq 0, \\
\left[\mu_i+\pi-\sigma_i,-\mu_i+2\pi-\sigma_i\right], &\text{if}\ X_{1i}<0,
\end{array}\right.
\end{split}
\end{align}
where $\mu_i\triangleq \arcsin(\frac{B_{2i}-B_{1i}}{\sqrt{X_{1i}^2+X_{2i}^2}})+2k\pi, k=\pm 1,\pm 2,\cdots$, and $\sigma_i\triangleq \arctan(\frac{X_{2i}}{X_{1i}})$. Here, $\Bbbk_i$ is cyclic and a complete cycle is $2\pi$. 

\item Case 2): If ${B_{2i}-B_{1i}}\leq -{\sqrt{X_{1i}^2+X_{2i}^2}}$, we have $\Bbbk_i=\mathbf{R}$, i.e., any $\varphi$ satisfies \eqref{eq-ineq}. 

\item Case 3): If ${B_{2i}-B_{1i}}\geq{\sqrt{X_{1i}^2+X_{2i}^2}}$, we have $\Bbbk_i\in\varnothing$, i.e., no feasible $\varphi$ can be found at the required ratio $C_{s_i}$. This case needs to be avoided by carefully configuring the values of $C_{s_i}$.
\end{itemize}

After obtaining the sets of all inequality constraints, we can derive the final range of $\varphi$ by finding their intersection. \cb{To make the comparison simpler, for each $\Bbbk_i$ satisfying Case 1), we select a segment in a $2\pi$-length section $[x,x+2\pi]$, where $x$ can be any real number. Let $x=-\pi$, i.e., the $2\pi$-length section is $[-\pi,\pi]$, and the selected segment is
\begin{align*}
\bar{\Bbbk}_i
&=\left\{
\begin{array}{cc}
[-\pi,\bar{\varphi}_{i2}]\cup[\bar{\varphi}_{i1},\pi], & \text{if } \pm\pi\in \Bbbk_i, \\ 
\left[\bar{\varphi}_{i1},\bar{\varphi}_{i2}\right], & \text{otherwise.}
\end{array}\right.
\end{align*}
Then we can obtain the intersection over the $2\pi$ period as   

\begin{align}
\varphi\in\bar{\mathcal{K}}&\triangleq\{\bar{\Bbbk}_1\cap\bar{\Bbbk}_2\cap\cdots\cap\bar{\Bbbk}_{N_s}\}=[\bar{\varphi}_{s1},\bar{\varphi}_{s2}].
\end{align}
}

It is worth noting that generally, the range of $\varphi$ decreases with the increase of $N_s$. For a specific constraint ${|\af^{T}(\theta_{s_i})\wf_t|}/{||\wf_t||}\geq C_{s_i}\sqrt{(1-\rho)M}$, reducing the value of $C_{s_i}$ can make $\varphi$ less constrained, and decrease the minimum gain in the direction $\theta_{s_i}$. \cb{Overall, the chance of $\bar{\mathcal{K}} = \varnothing$ grows with the increase of $N_s$ and $C_{s_i}$. This extreme case happens when the constraints lead to an empty intersection of $\varphi$. In this case, we can use two possible ways to obtain alternative, suboptimal solutions of $\bar{\mathcal{K}}$. One is to partially relax the constraints, by progressively reducing the value of $C_{s_i}$ or discarding part of the constraints until $\bar{\mathcal{K}}\neq\varnothing$. The other is to firstly obtain the \textit{main segment} of $\varphi$ as the section $[\varphi_{1\text{m}},\varphi_{2\text{m}}]$, and then refine it by considering the constraint(s) prioritized to be met first. The constraints can be, but are not limited to, the one constraining BF gain in the dominating AoD.

The feasible range of $\varphi$ can be then expressed as} 
\begin{align}
\varphi\in\Bbbk_s=[\varphi_{s1},\varphi_{s2}]=[\bar{\varphi}_{s1}+&2k\pi,\bar{\varphi}_{s2}+2k\pi],\\ \notag
&k=\pm 1,\pm 2,\cdots.
\end{align}

After that, by comparing $\Bbbk_s$ with $\varphi_\text{opt}$, the constrained optimal combining phase can be obtained as  
\cb{
\begin{align}\label{eq-phiopt}
\begin{split}
{\varphi}^{(1)}_\text{opt}=\left\{\begin{array}{cc}
\varphi_\text{opt}, &\text{if}\ \varphi_\text{opt}\in \Bbbk_s, \\
\varphi_{s1}, & \text{if}\ \varphi_\text{opt}\notin \Bbbk_s \text{ and } f(\varphi_{s1}) \leq f(\varphi_{s2}),\\
\varphi_{s2}, & \text{if}\ \varphi_\text{opt}\notin \Bbbk_s \text{ and } f(\varphi_{s1})  > f(\varphi_{s2}),
\end{array}\right.
\end{split}
\end{align}
where $f(\varphi)={\wtf^H\Ht^H\Ht\wtf}/{\|\wtf\|^2}$. Since the period of $f(\varphi)$ is $2\pi$, and the range $[\varphi_{s1},\varphi_{s2}]$ is no greater than $2\pi$, referring to the monotonicity analysis in Appendix \ref{apdx-monotonicity}, we can see that when $\varphi_\text{opt} \notin \mathcal{K}$, the optimal value $\varphi_\text{opt}^{(1)}$ is reached at either $\varphi_{s1}$ or $\varphi_{s2}$. By comparing the values of $f(\varphi_{s1})$ and $f(\varphi_{s2})$, we can determine the optimal solution, as described in \eqref{eq-phiopt}.}

According to \cite{luo2019optimization}, the complexity of calculating $\varphi_\text{opt}$ is $O(M^2)$, and the additional complexity of calculating ${\varphi}^{(1)}_\text{opt}$ is also bounded by $O(M^2)$, since $N_s\leq M$ in most cases.

When a single constraint on the desired scanning direction to which $\wf_{t,s}$ points is employed, a relatively simple and practical solution can be obtained without looking into complicated computation of the intersection. In this case, $\Bbbk_s=\Bbbk_1=[\varphi_{11},\varphi_{21}]$, where $\Bbbk_1$ is the range of $\varphi$. $\varphi_\text{opt}^{(1)}$ can be then obtained as 
\cb{
\begin{align}\label{eq-phiopt1}
\begin{split}
{\varphi}^{(1)}_\text{opt}=\left\{\begin{array}{cc}
\varphi_\text{opt}, &\text{if}\ \varphi_\text{opt}\in \Bbbk_s, \\
\varphi_{11}, & \text{if}\ \varphi_\text{opt}\notin \Bbbk_s \text{ and } f(\varphi_{11}) \leq f(\varphi_{12}),\\
\varphi_{12}, & \text{if}\ \varphi_\text{opt}\notin \Bbbk_s \text{ and } f(\varphi_{11})  > f(\varphi_{12}).
\end{array}\right.
\end{split}
\end{align}	
}

\subsubsection{Constrained Total Scanning Power over a Range of Directions}\label{sec-ConsPow}

As shown in Section \ref{sec-ScanGain}, when $N_s$ is large, finding the range for $\varphi$ that meets the gain constraints on multiple discrete directions can be operationally complicated. More practically, we can set a minimum total power constraint over a range of scanning directions. In this section, we investigate the optimization problem under such a minimum total power constraint. The problem can be formulated as
\cb{
\begin{subequations}
	\begin{align}
	&\text{P}_2:\  {\varphi}^{(2)}_\text{opt}=\arg\max_{\varphi}\dfrac{\wtf^H\Ht^H\Ht\wtf}{\|\wtf\|^2}, \label{eq-Hcase3}\\
	&\text{s.t. }\ \int_{\theta_{s_l}}^{\theta_{s_r}}\dfrac{|\af^{T}(\theta)\wf_t|^2}{||\wf_t||^2}d\theta \geq C_{sp}\int_{\theta_{s_l}}^{\theta_{s_r}}|\af^{T}(\theta)\wf_2|^2d\theta, \label{eq-stScanPow}\\
	&\text{with}\ \mathbf{w}_t=\sqrt{\rho}\wcf+\sqrt{1-\rho} e^{j\varphi}\wsf. \notag
	\end{align}
\end{subequations}
The integrand ${|\af^{T}(\theta)\wf_t|^2}/{||\wf_t||^2}$ on the left-hand side of \eqref{eq-stScanPow} is the normalized BF gain in the direction of $\theta$, and $\theta_{s_l}$ and $\theta_{s_r}$ are the bounds of the BF range of interest. On the right-hand side of \eqref{eq-stScanPow}, $C_{sp}$ is a scaling coefficient; and $\wf_2$, $\|\wf_2\|=1$, is the BF weight optimized for the BF waveform in Method 2 in \cite{zhang2019multibeam}.} The threshold does not affect our methodology for solving this problem and can change to different values. We use the one in \eqref{eq-stScanPow} to provide a concrete reference only.  

Note that the integration is conducted based on $\theta$ and is independent of $\wf_t$, we can move $\wf_t$ out of the integration in \eqref{eq-stScanPow}. This leads to
\begin{align}
\int_{\theta_{s_l}}^{\theta_{s_r}}\dfrac{|\af^{T}(\theta)\wf_t|^2}{||\wf_t||^2}d\theta =\dfrac{\wf_t^H\left(\int_{\theta_{s_l}}^{\theta_{s_r}}\At_{int}(\theta) \ d \theta  \right)\wf_t}{||\wf_t||^2},
\label{eq-inttheta}
\end{align}
$\At_{int}(\theta)=\af^{*}(\theta)\af^{T}(\theta)$. The integration on the right-hand side of \eqref{eq-inttheta} is based on each element in the matrix $\At_{int}(\theta)$, and the output of the integration is also a matrix.

We cannot obtain a closed-form result for the integral of each element in $\At_{int}(\theta)$

We can instead approximate the integral as a summation, as follows. 
\begin{align}\label{eq-integral}
\bm{\mathcal{A}}=\int_{\theta_{s_l}}^{\theta_{s_r}}\At_{int}(\theta)d\theta\approx\sum^{N_I}_{i=1}\delta_\theta\At_{int}(\theta_{s_1}+i\delta_\theta),
\end{align}
where $\delta_\theta=(\theta_{s_r}-\theta_{s_l})/N_I$ is the step size and $N_I$ is the total number of steps. It is assumed that $N_I$ is large enough to guarantee a small enough step size. \cb{We use the primitive form of numerical integration because the elements in $\bm{\mathcal{A}}$ have complex values, which makes it hard to implement numerical integration algorithms developed mostly in the real space.} For a set of values of $\theta_{s_l}$ and $\theta_{s_r}$, we can pre-calculate and store the numerical results. Since $\At_{int}$ is a Toeplitz matrix, only $(2M-1)$ numerical integrations are calculated and stored for a given range of sensing BF directions. The complexity of calculating $\bm{\mathcal{A}}$ is $O(MN_I)$. 

Once the matrix $\bm{\mathcal{A}}$ is obtained, we can proceed to specify the range of $\varphi$, i.e., $[\varphi_{p1},\varphi_{p2}]$, according to the constraint. The derivation process is similar to that in Section \ref{sec-ScanGain}, and provided together with the results in Appendix A.  

Therefore, $ {\varphi}^{(2)}_\text{opt}$ under the constrained total power can be obtained as
\cb{
\begin{align}\label{eq-phiopt2}
\begin{split}
{\varphi}^{(2)}_\text{opt}=\left\{\begin{array}{cc}
\varphi_\text{opt}, &\text{if}\ \varphi_\text{opt}\in \Bbbk_p, \\
\varphi_{p1}, & \text{if}\ \varphi_\text{opt}\notin \Bbbk_p \text{ and } f(\varphi_{p1}) \leq f(\varphi_{p2}),\\
\varphi_{p2}, & \text{if}\ \varphi_\text{opt}\notin \Bbbk_p \text{ and } f(\varphi_{p1})  > f(\varphi_{p2}),
\end{array}\right.
\end{split}
\end{align}
}

The complexity of calculating $ {\varphi}^{(2)}_\text{opt}$ is $O(\max{\{M^2,MN_I\}})$.

\subsection{Optimizing Scanning Subbeam with Constraint on Received Signal Power}\label{sec-OPTscan}

We can also optimize the BF waveform of the scanning subbeam while meeting the constraint on the received signal power for communications. The optimization problem is formulated as
\cb{\begin{subequations}
 	\begin{align}
	&\text{P}_3:\  {\varphi}^{(3)}_\text{opt}=\arg\max_{\varphi}\dfrac{|\af^T(\theta_{s_0})\wtf|^2}{\|\wtf\|^2}, \label{eq-MaxScan}\\
	\text{or } &\text{P}_4:\   {\varphi}^{(4)}_\text{opt}=\arg\max_{\varphi}\int_{\theta_{s_l}}^{\theta_{s_r}}\dfrac{|\af^{T}(\theta)\wf_t|^2}{||\wf_t||^2}d\theta, \label{eq-MaxSPow}\\
	&\text{s.t.}\ \ \ \dfrac{\wtf^H\Ht^H\Ht\wtf}{||\wf_t||^2}\geq C_{p} P_c, \label{eq-stOutPow}\\	
	&\text{with}\ \mathbf{w}_t=\sqrt{\rho}\wcf+\sqrt{1-\rho} e^{j\varphi}\wsf.\notag 
	\end{align} 
\end{subequations}}where \eqref{eq-MaxScan} maximizes the gain at the dominating AoD of the scanning beam, and \eqref{eq-MaxSPow} maximizes the power over a range of scanning directions. $C_p$ is the scaling coefficient, and {$P_c=\|\Ht\wcf\|^2$} is the output signal power when only a single communication beam is used. Here, either \eqref{eq-MaxScan} or \eqref{eq-MaxSPow} is used, depending on the objective. For this optimization problem, we can first find the optimal solution to one of \eqref{eq-MaxScan} and \eqref{eq-MaxSPow}, and then check it against the range that can be obtained from \eqref{eq-stOutPow}. 

Similar to \eqref{eq-ineq}, the objective function of \eqref{eq-MaxScan} can be rewritten as
\begin{align*}
g(\varphi)=\frac{[\rho b_{20}^2+(1-\rho)b_{30}^2]+2Pb_{20}b_{30}\cos (\varphi+\beta_{20}+\beta_{30})}{1+2Pb_1\cos(\varphi+\beta_1)},
\end{align*}
where  $\wcf^H\wsf=b_1e^{j\beta_1}$, $\wcf^H\af^*(\theta_{s_0})=b_{20}e^{j\beta_{20}}$, and $\af^T(\theta_{s_0})\wsf=b_{30}e^{j\beta_{30}}$. By letting $g'(\varphi)=0$ and analyzing the monotonicity of $g(\varphi)$, we can obtain $\varphi_\text{smax}$, which achieves the maximal value of $g(\varphi)$, as
\begin{align}
\varphi_{\text{smax}}^{(3)}=\left\{
\begin{array}{cc}
\pi+\eta_0-\zeta_0+2k\pi,& \text{if} \ D_1\geq 0,\\
\eta_0-\zeta_0+2k\pi, &  \text{if} \ D_1<0,
\end{array}\right. \notag\\
k=0,\pm1,\pm2\cdots
\end{align}
where
\begin{align}\label{eq-interVar2}
\begin{split}
\eta_0\triangleq&\arcsin{(L_s/\sqrt{D_1^2+D_2^2})},\ \zeta_0\triangleq\arctan(D_2/D_1),\\
D_1\triangleq&-2Pb_{20}b_{30}\cos(\beta_2+\beta_3)\\
&+2Pb_1[\rho b_{20}^2+(1-\rho)b_{30}^2]\cos\beta_1\\
D_2\triangleq&-2Pb_{20}b_{30}\sin(\beta_2+\beta_3)\\
&+2Pb_1[\rho b_{20}^2+(1-\rho)b_{30}^2]\sin\beta_1\\
L_s\triangleq&-4P^2b_1b_{20}b_{30}\sin(\beta_2+\beta_3-\beta_1).\\
\end{split}
\end{align} 
Similarly, if \eqref{eq-MaxSPow} is used as the objective function, we can obtain
\begin{align}
\varphi_{\text{smax}}^{(4)}=\left\{
\begin{array}{cc}
\pi+\etat_0-\tilde{\zeta}_0+2k\pi,& \text{if} \ \tilde{D}_1\geq 0,\\
\etat_0-\tilde{\zeta}_0+2k\pi, &  \text{if} \ \tilde{D}_1<0,
\end{array}\right. \notag\\
k=0,\pm1,\pm2\cdots
\end{align}
where
\begin{align}
\begin{split}
\etat_0\triangleq&\arcsin{\Big(\Lt_s/\sqrt{\tilde{D}_1^2+\tilde{D}_2^2}\Big)},\ \tilde{\zeta}_0\triangleq\arctan(\tilde{D}_2/\tilde{D}_1),\\ 
\tilde{D}_1\triangleq&-2P|b_p|\cos\beta_p+\\
& 2P|b_1|[\rho\wcf^H\bm{\mathcal{A}}\wcf+(1-\rho)\wsf^H\bm{\mathcal{A}}\wsf]\cos\beta_1,\\
\tilde{D}_2\triangleq&-2P|b_p|\sin\beta_p+\notag\\
&2P|b_1|[\rho\wcf^H\bm{\mathcal{A}}\wcf+(1-\rho)\wsf^H\bm{\mathcal{A}}\wsf]\sin\beta_1,\\
\tilde{L}_s\triangleq&4P^2|b_1||b_p|\sin(\beta_1-\beta_p).\\
\end{split}
\end{align}

The range of $\varphi$ determined by \eqref{eq-stOutPow}, can be derived in a similar way to \eqref{eq-stscan} and \eqref{eq-stScanPow}, and the detail is provided in Appendix \ref{apdx-phig}. With the range of $\varphi$, $[\varphi_{g_1},\varphi_{g_2}]$ given in Appendix \ref{apdx-phig}, the optimal solutions, $\varphi_\text{opt}^{(3)}$ and $\varphi_\text{opt}^{(4)}$, to problems $\text{P}_3$ and $\text{P}_4$ can be obtained as 
\cb{	
\begin{align}\label{eq-phiopt3}
\begin{split}
{\varphi}_\text{opt}^{(3)}=\left\{\begin{array}{cc}
\varphi_\text{smax}^{(3)}, &\text{if}\ \varphi_\text{smax}^{(3)}\in \Bbbk_g, \\
\varphi_{g1}, & \text{if}\ \varphi_\text{smax}^{(3)}\notin \Bbbk_g \text{ and } g(\varphi_{g1}) \leq g(\varphi_{g2}),\\
\varphi_{g2}, & \text{if}\ \varphi_\text{smax}^{(3)}\notin \Bbbk_g \text{ and } g(\varphi_{g1})  > g(\varphi_{g2}),
\end{array}\right.
\end{split}
\end{align}
}
or
\cb{
\begin{align}
\begin{split}
{\varphi}_\text{opt}^{(4)}=\left\{\begin{array}{cc}
{\varphi}_{\text{smax}}^{(4)}, &\text{if}\ {\varphi}_{\text{smax}}^{(4)}\in \Bbbk_g, \\
\varphi_{g1}, & \text{if}\ {\varphi}_{\text{smax}}^{(4)}\notin \Bbbk_g \text{ and } g(\varphi_{g1}) \leq g(\varphi_{g2}),\\
\varphi_{g2}, & \text{if}\ {\varphi}_{\text{smax}}^{(4)}\notin \Bbbk_g \text{ and } g(\varphi_{g1})  > g(\varphi_{g2}),
\end{array}\right.\\
\end{split}
\end{align}
}

To calculate ${\varphi}_\text{opt}^{(3)}$, the complexity is upper bounded by $O(M^2)$. Similar to the optimization in Section \ref{sec-ConsPow}, the complexity of calculating ${\varphi}_\text{opt}^{(4)}$ is $O(\max{\{M^2,MN_I\}})$.

\cb{The above \textit{subbeam-combiner} methods have a quadratic complexity ($O(M^2)$ or $O(MN_I)$). In real-time operations, $\wtf$ is regenerated every packet. Even with tens of antenna elements, e.g., $M=64$, the period is long enough for the calculations described above with the state-of-the-art commercial signal processing devices, such as Intel$^\circledR$ Stratix$^\circledR$ 10 FPGAs\cite{stratix10}.}

\section{Global Optimization Using SDR}\label{sec-SDR}
The optimization methods proposed in Section \ref{sec-OptPhi}, as well as those in \cite{zhang2019multibeam} and \cite{luo2019optimization}, seek the optimal combining weight $\varphi$ based on the pre-generated, known BF vectors $\wcf$ and $\wsf$. These results are relatively simple and practical for implementation, but they are sub-optimal. In this section, we develop global optimization methods that directly optimize $\wtf$, considering communication and sensing requirements. These methods allow us to obtain near-optimal solutions, and enable us to evaluate the performance loss of the suboptimal solutions. We first study the constrained maximization of the received signal power for communication, and then the constrained optimization of the BF waveform for sensing.

\subsection{Maximizing Received Signal Power with Constraints on BF Waveform}\label{sec-MaxPow}
We first maximize the received signal power for communication, with one or other constraints on the BF waveform. The global optimization problem for $\wtf$ can be formulated as
\begin{subequations}\label{eq-formulation2}
	\begin{align} 
	&\text{P}_5:\ \wtf^{(5)}= \argmax_{\wtf, \wtf^H\wtf=1} \ \wtf^H\Ht^H\Ht\wtf, \label{eq-OutPow2}\\
	&\text{s.t.}\quad\|\Dt (\At\wtf-c_s\df_v)\|^2\leq \varepsilon_w, \label{eq-waveform2}\\
	&\qquad\ |\af(\theta_{s_i})^T\wtf|^2\geq \varepsilon_{s_i},\ i=1,2,\cdots, N_s, \ \text{and/or} \label{eq-AoDS2}\\
	&\qquad\int_{\theta_{s_l}}^{\theta_{s_r}}|\af(\theta)^T\wtf|^2d\theta\geq\varepsilon_p, \label{eq-intPow2}, 
	\end{align}
\end{subequations}
where \eqref{eq-waveform2} bounds the mismatches between the generated and the desired BF waveforms (i.e., array radiation patterns), \eqref{eq-AoDS2} constrains the gain of the scanning subbeam in $N_s$ concerned directions, \eqref{eq-intPow2} constrains the power over a range of consecutive scanning directions, and $\varepsilon_w$, $\varepsilon_{s_i}$ and $\varepsilon_p$ are the bounds for those constraints. These constraints can be applied individually or jointly. The constraints \eqref{eq-AoDS2} and \eqref{eq-intPow2} correspond to those we have discussed in Section \ref{sec-OptPhi}. We elaborate on the constraint \eqref{eq-waveform2} below. 

In \eqref{eq-waveform2}, $\At=[\af(\theta_1),\af(\theta_2),\cdots,\af(\theta_N)]^T$ is the array response matrix in $N$ specified directions. $\Dt$ is a pre-chosen  diagonal weighting matrix that can be used to impose different accuracy requirements on different segments of the generated BF waveform. $c_s$ is a real scaling factor. $\df_v=\Dt_v\pf_v$ is the complex desired BF waveform, where $\Dt_v$ is a diagonal matrix with diagonal elements being the magnitude of the desired BF waveform, and $\pf_v$ is a vector containing the corresponding phase. In most cases, only $\Dt_v$ needs to be specified, and $\pf_v$ can be optimized by using, e.g., a two-step ILS method \cite{shi2005new}. {When ILS is used, the value of $\pf_v$ is updated by $\pf_v^{(l)}=\exp\{j\arg(\At\wtf^{(l-1)})\}$ at the $l$th literation.} The scaling factor $c_s$ can be determined to minimize the BF waveform mismatch. Taking the derivative of $\|\Dt (\At\wtf-c_s\df_v)\|^2$ with respect to $c_s$ and letting it be zero, we can obtain 
\begin{align}\label{eq-cs0}
c_{s}=\frac{\mathfrak{Re}\{\df_v^H\Dt^H\Dt\At\wtf\}}{\|\df_v\|^2}.
\end{align}
Then we can rewrite \eqref{eq-waveform2} as
\begin{align}\label{eq-WVFM}
\|\Dt\At\wtf\|^2-{\mathfrak{Re}^2\{\df_v^H\Dt^H\Dt\At\wtf\}}/{\|\df_v\|^2}\leq \varepsilon_w.
\end{align}

Since \eqref{eq-formulation2} is a nonconvex NP-hard problem, it is challenging to obtain a closed-form solution to $\wtf$. However, we can convert this problem to a homogeneous QCQP problem and apply the SDR technique \cite{ma2010semidefinite}.

\cb{We first reformulate the original complex optimization problem to a real one, because \eqref{eq-waveform2} cannot be directly converted to the standard form of a complex semidefinite programming (SDP) constraint.} Let
\begin{align}\label{eq-real}
\begin{split}
&\tilde{\At}\triangleq\begin{bmatrix}&\mathfrak{Re}\{\At\} &-\mathfrak{Im}\{\At\}\\
&\mathfrak{Im}\{\At\} &\mathfrak{Re}\{\At\}
\end{bmatrix},\\
&\tilde{\bm{\mathcal{A}}}\triangleq\begin{bmatrix}&\mathfrak{Re}\{\bm{\mathcal{A}}\} &-\mathfrak{Im}\{\bm{\mathcal{A}}\}\\
&\mathfrak{Im}\{\bm{\mathcal{A}}\} &\mathfrak{Re}\{\bm{\mathcal{A}}\}
\end{bmatrix},\\
&\tilde{\Ht}\triangleq\begin{bmatrix}&\mathfrak{Re}\{\Ht\} &-\mathfrak{Im}\{\Ht\}\\
&\mathfrak{Im}\{\Ht\} &\mathfrak{Re}\{\Ht\}
\end{bmatrix},\ \tilde{\Dt}\triangleq\begin{bmatrix}&\Dt &\bm{0}\\
&\bm{0} &\Dt
\end{bmatrix},\\
&\tilde{\At}_{s_i}\triangleq\begin{bmatrix}&\mathfrak{Re}\{\af^T(\theta_{s_i})\}&-\mathfrak{Im}\{\af^T(\theta_{s_i})\}\\&\mathfrak{Im}\{\af^T(\theta_{s_i})\}&\mathfrak{Re}\{\af^T(\theta_{s_i})\}\end{bmatrix},\\
&\tilde{\wf}_t\triangleq\begin{bmatrix}\mathfrak{Re}\{\wtf^T\}\ \ \mathfrak{Im}\{\wtf^T\}\end{bmatrix}^T,\\ &\tilde{\df}_v\triangleq\begin{bmatrix}\mathfrak{Re}\{\df_v^T\}\ \ \mathfrak{Im}\{\df_v^T\}\end{bmatrix}^T,
\end{split}
\end{align}
where $\tilde{\wf}_t$ and $\tilde{\df}_v$ are $2M\times 1$ vectors, and $\tilde{\At},\tilde{\bm{\mathcal{A}}},\tilde{\At}_{s},\tilde{\Ht},\tilde{\Dt}\in\mathbb{S}^{2M}$. As shown in Appendix \ref{apdx-Complex2Real}, using these real variables, we can recast the problem with complex variables, \eqref{eq-formulation2}, to 

\begin{subequations}\label{eq-realPro2}
\begin{align}
&\wtft^{(5)}=\argmin_{\wtft, \tilde{\wf}_t^T\tilde{\wf}_t=1}-\tilde{\wf}_t^T\tilde{\Ht}^T\tilde{\Ht}\tilde{\wf}_t ,\\
&\text{s.t.}\quad \tilde{\wf}_t^T\tilde{\At}^T\tilde{\Dt}^T(\It-\frac{\tilde{\Dt}\tilde{\df}_v\tilde{\df}_v^T\tilde{\Dt}^T}{\|\tilde{\df}_v\|^2})\tilde{\Dt}\tilde{\At}\tilde{\wf}_t \leq \varepsilon_w, \label{eq-realConsWVFM}\\
&\qquad\tilde{\wf}_t^T\tilde{\At}_{s_i}^T\tilde{\At}_{s_i}\tilde{\wf}_t\geq\varepsilon_{s_i},\ i=1,2,\cdots, N_s \ \text{and/or}\\
&\qquad\ \tilde{\wf}_t^T\tilde{\bm{\mathcal{A}}}\tilde{\wf}_t\geq\varepsilon_p.
\end{align}
\end{subequations}
where $\It$ is a $2M\times2M$ identity matrix. \eqref{eq-realPro2} is an inhomogeneous QCQP problem, and can be further converted to a homogeneous QCQP problem. Let
\begin{align}\label{eq-hat1}
\begin{split}
&\hat{\At}\triangleq \tilde{\At}^T\tilde{\Dt}^T(\It-\tilde{\df}_v\tilde{\df}_v^T/\|\tilde{\df}_v\|^2)\tilde{\Dt}\tilde{\At},\\ 
&\hat{\At}_{s_i}\triangleq \tilde{\At}_{s_i}^T\tilde{\At}_{s_i},\ \hat{\Ht}\triangleq\tilde{\Ht}^T\tilde{\Ht},
\end{split}
\end{align} 
where $\hat{\At},\hat{\At}_{s_i},\hat{\Ht}\in\mathbb{S}^{2M}$. We can rewrite \eqref{eq-realPro2} as
\begin{align}\label{eq-homoPro2}
\begin{split}
\wtft^{(5)}&=\argmin_{\wtft, \wtft^T\wtft=1}\ \wtft^T\hat{\Ht}\wtft,\\
\text{s.t.}\ &\ \wtft^T\hat{\At}\wtft\leq \varepsilon_w,\\
&\ \wtft^T\hat{\At}_{s_i}\wtft\geq\varepsilon_{s_i},\ i=1,2,\cdots, N_s,\  \text{and/or}\\ 
&\ \wtft\tilde{\bm{\mathcal{A}}}\wtft\geq\varepsilon_p.
\end{split}
\end{align}
The real-valued homogeneous QCQP problem in \eqref{eq-homoPro2} can be relaxed to
\begin{align}\label{eq-SDP2}
\begin{split}
\Wt^{(5)}=&\argmin_{\Wt,  \Tr {(\Wt)}=1,\ \Wt\geq0}\Tr {(-\hat{\Ht}\Wt)}\\
\text{s.t. }\ &\Tr {(\hat{\At}\Wt)}\leq \varepsilon_w,\\
&\Tr {(\hat{\At}_{s_i}\Wt)}\geq\varepsilon_{s_i}, \ i=1,2,\cdots, N_s, \ \text{and/or}\\
 &\Tr {(\tilde{\bm{\mathcal{A}}}\Wt)}\geq\varepsilon_p,
\end{split}
\end{align}
where $\Wt=\tilde{\wf}_t\tilde{\wf}_t^T$, and $\Wt\geq0$ indicates that $\Wt$ is positive definite. 

The problem \eqref{eq-SDP2} can now be solved by SDP from the standard convex optimization toolbox CVX \cite{grant2008cvx}, and the globally optimal solution $\Wt^{(5)}$ to \eqref{eq-SDP2} can be obtained. Once $\Wt^{(5)}$ is obtained, we can apply several different ways \cite{ma2010semidefinite} to obtain an approximated solution for $\wtft^{(5)}$, and the simplest one is to apply the eigen-decomposition to $\Wt^{(5)}$. \cb{The eigen-decomposition method is efficient since the rank of $\Wt^{(5)}$ is observed to be very low (mostly 1 and occasionally 2) in our simulations. This also applies to the following three problem formulations in this section.} Let
\begin{align}\label{eq-eigen}
\tilde{\wf}_t^\star=\sqrt{\lambda_1}\qf_1,
\end{align}
where $\lambda_1$ is the maximal eigenvalue of $\Wt^{(5)}$, and $\qf_1$ is its corresponding eigenvector.
Then the complex BF vector $\wtf^\star$ is given by
\cb{
\begin{align}\label{eq-finalw}
\wtf^\star=\frac{\wtft^\star{[1:M]}+j\,\wtft^\star{[M+1:2M]}}{||\tilde{\wf}_t||^2},
\end{align}}where $\wtft^\star{[1:M]}$ and $\wtft^\star{[M+1:2M]}$ denote the first and last $M$ elements of $\wtft$, respectively. \cb{The normalization is applied to $\wtf^\star$ to make the power of $\wtf^\star$ equal to 1.}

Since the optimal $\pf_v$ cannot be directly obtained in one iteration, the computation is recursively applied several times until convergence or the maximal number of iterations is reached. The iterative algorithm is summarized in Algorithm \ref{alg-SDR-ILS}. Similar to the ILS approach in \cite{shi2005new}, the suboptimal value of $\pf_v$ can be iteratively calculated, and the optimization algorithm can be shown to converge after a few iterations in most cases. 
\begin{algorithm}[tb]
	\caption{SDP-ILS Algorithm}\label{alg-SDR-ILS}
	\textbf{Input}: $\Ht$, $\At$, $\bm{\mathcal{A}}$, $\Dt_v$, $\pf_{v_0}=[1,\cdots,1]^T$, $\theta_s$, $L_{\text{max}}$, $\varepsilon_{s_i}$, $\varepsilon_{p}$, $\varepsilon_w$.\\
	\textbf{Output}: Global optimized $\wtf^{(5)}$, $\gamma_{max}$\\
	\textbf{0)} $\pf_v=\pf_{v_0}$, go to 1);\\
	\textbf{1)} If $l<L_\text{max}$, let $\df_v=\Dt_v\pf_v$; and compute $\hat{\Ht}$, $\hat{\At}$, $\hat{\At}_{s_i}$, $\tilde{\bm{\mathcal{A}}}$ through \eqref{eq-real} and \eqref{eq-hat1}, go to 2); If $l=L_\text{max}$, go to 5);\\
	\textbf{2)} Compute $\Wt^{(5)}$ in \eqref{eq-SDP2} using SDP, go to 3); \\
	\textbf{3)}	Calculate the approximate $\wtf^\star$ using \eqref{eq-eigen} and \eqref{eq-finalw}, or other methods, e.g., the randomization procedure in \cite{ma2010semidefinite}. Go to 4); \\
	\textbf{4)} With $\wtf^\star$, let $\pf_v=\exp\{j\arg(\At\wtf^\star)\}$, go to 1);\\
	\textbf{5)} Let $\wtf^{(5)}=\wtf^\star$, and compute the maximal received signal power by $\gamma_{max}=\|\Ht\wtf^{(5)}\|^2$. 

\end{algorithm} 

\subsection{Constrained Optimization of BF Waveform} \label{sec-OptWvfm}
Seeking the global optimal solutions, we can also target at optimizing the BF waveform of the scanning subbeam under various constraints. Such an optimization problem can be formulated in different ways. Here, we consider an example of minimizing the mismatch between the desired and the generated BF waveforms. The problem can be formulated as
\begin{subequations}\label{eq-formulation1}
\begin{align} 
\text{P}_6:\ &\wtf^{(6)}=\argmin_{\wtf,\wtf^H\wtf=1,c_s,\pf_v}\|\Dt (\At\wtf-c_s\Dt_v\pf_v)\|^2, \label{eq-waveform1}\\
\text{s.t.}&\ \wtf^H\Ht^H\Ht\wtf\geq C_{p} P_c, \label{eq-OutPow1}\\
&\ |\af(\theta_{s_i})^T\wtf|^2\geq \varepsilon_{s_i},\ i=1,2,\cdots, N_s\ \text{(optional)},\label{eq-AoDS1} \\
&\ \int_{\theta_{s_l}}^{\theta_{s_r}}|\af(\theta)^T\wtf|^2d\theta\geq\varepsilon_p\ \text{(optional)}, \label{eq-intPow1}
\end{align}
\end{subequations}
where \eqref{eq-OutPow1} requires the received signal power to meet the communication requirement, and the other two constraints \eqref{eq-AoDS1} and \eqref{eq-intPow1} are optional. Using the value of $c_{s}$ in \eqref{eq-cs0}, we can rewrite the objective function \eqref{eq-waveform1} as the left-hand side of inequality \eqref{eq-WVFM}.

Similar to the derivation in Section \ref{sec-MaxPow}, we can relax the original problem with complex variables in \eqref{eq-formulation1} to a homogeneous QCQP problem with real variables. The relaxed version of \eqref{eq-formulation1} can be obtained as
\begin{align}\label{eq-SDP1}
\begin{split}
\Wt^{(6)}=&\argmin_{\Wt,\Tr {(\Wt)}=1,\ \Wt\geq0}\Tr {(\hat{\At}\Wt)}\\
\text{s.t. }&\Tr {(\hat{\Ht}\Wt)}\geq C_{p} P_c,\\
&\Tr {(\hat{\At}_{s_i}\Wt)}\geq\varepsilon_{s_i},\ i=1,2,\cdots, N_s\ \text{(optional)},\\
&\Tr {(\tilde{\bm{\mathcal{A}}}\Wt)}\geq\varepsilon_p\ \text{(optional)},
\end{split}
\end{align}
which can be solved by SDP. A suboptimal solution for $\wtf^\star$ can be obtained. 

We may formulate other objective functions, such as maximizing the BF gain over some specified directions or energy over a range of directions, subject to constraints on communication performance. These problems can be solved in the same way as in the above example. Here, we only list two alternative formulations that will be simulated for comparison with other schemes. The details are omitted.
\begin{enumerate}
	\item Maximize BF gain in specified directions: 
\begin{align}\label{eq-Globe3}
\begin{split}
&\text{P}_7:\ \wtf^{(7)}= \argmax_{\wtf,\wtf^H\wtf=1} \ |\af(\theta_{s_0})^T\wtf|^2, \\
&\text{s.t.}\quad\|\Dt (\At\wtf-c_s\df_v)\|^2\leq \varepsilon_w, \\
&\qquad\ \wtf^H\Ht^H\Ht\wtf\geq C_{p} P_c,\\
&\quad\text{optionally, }\\
&\qquad|\af(\theta_{s_i})^T\wtf|^2\geq \varepsilon_{s_i}\ i=1,2,\cdots, N_s-1; 
\end{split}
\end{align}

\item Maximize BF energy over a given range:
{
\begin{align}\label{eq-Globe4}
\begin{split}
&\text{P}_8:\ \wtf^{(8)}= \arg \max_{\wtf} \int_{\theta_{s_l}}^{\theta_{s_r}}|\af(\theta)^T\wtf|^2d\theta,\\ 
&\text{s.t.}\quad\|\Dt (\At\wtf-c_s\df_v)\|^2\leq \varepsilon_w, \\
&\qquad\ \wtf^H\Ht^H\Ht\wtf\geq C_{p} P_c,\\
&\quad\text{optionally, }\\
&\qquad|\af(\theta_{s_i})^T\wtf|^2\geq \varepsilon_{s_i},\ i=1,2,\cdots, N_s.
\end{split} 
\end{align}
}
\end{enumerate}

\subsection{Complexity of Global Optimization}

The complexity of the proposed global optimization methods is much higher than the sub-optimal solutions developed in Section \ref{sec-OptPhi}, due to the iterative use of SDP.

Applying SDP, the proposed methods in Section \ref{sec-SDR} have polynomial complexities. In the worst case, the complexity is $O(L_\text{max}\text{max}\{N_{cs},2M\}^4\sqrt{2M}\log(1/\epsilon))$, where $\epsilon>0$ is the required solution accuracy and $N_{cs}$ is the number of constraints, e.g., $N_{cs}=N_s+3$ if \eqref{eq-SDP2} is solved. For the JCAS system, $2M$ is typically greater than $N_{cs}$. Hence the worst-case computational complexity is $O(16\sqrt{2}L_\text{max}M^{4.5}\log(1/\epsilon))$. As will be observed in simulations, the algorithms can typically converge within 3 to 6 iterations. The complexity can be reduced by employing fast real-time convex optimization solvers which use the possible special features of the data matrices' structures such as sparsity \cite{ma2010semidefinite,mattingley2010real}. Usually, the computational complexity practically achieved by the SDP solvers \cite{grant2008cvx} is much lower than the worst-case complexity. 

\cb{For arrays with medium numbers of antenna elements, e.g., $M=10$, even with the worst-case complexity, the real-time implementation of our algorithms is possible, with advanced commercial signal processing hardware possessing computing performance of more than ten teraFLOPS (TFLOPS), such as Intel$^\circledR$ Agilex$^\text{TM}$ \cite{agilexRtable}. When the number of array elements is large, the implementation of the algorithms can be costly at present. The algorithms still provide benchmarks for performance evaluation of suboptimal solutions.}

\section{Simulation Results}\label{sec-simulation}
In this section, simulation results are presented to verify the proposed optimization methods. The proposed methods are compared to three existing schemes: Methods 1 and 2 in \cite{zhang2019multibeam}, and the method developed in \cite{luo2019optimization} without any constraint on sensing waveform, which are denoted by ``M1-Zhang", ``M2-Zhang" and ``Without Cons" in the legends of all the figures, respectively. "M2-Zhang" and "Without Cons" can be treated as the benchmark methods that achieve superior BF waveform and received signal power for communications, respectively. The solutions to the problem formulations $\text{P}_1$, $\text{P}_2$, $\text{P}_3$, and $\text{P}_4$ in Section \ref{sec-OPTscan} are denoted as ``$\text{P}_1$: RxP-SG", ``$\text{P}_2$: RxP-SP", ``$\text{P}_3$: SG-RxP", and ``$\text{P}_4$: SP-RxP ", respectively. The solutions to problems $\text{P}_5$, $\text{P}_6$, $\text{P}_7$, and $\text{P}_8$ in Section \ref{sec-SDR}  are denoted as ``$\text{P}_5$: SDP-RxP", ``$\text{P}_6$: SDP-Err", ``$\text{P}_7$: SDP-SG", and ``$\text{P}_8$: SDP-SP", respectively.

\subsection{Simulation Setup}
For all simulations, a ULA with $M=16$ omnidirectional antennas (spaced at half wavelength) is used. We assume that the basic reference fixed subbeam points at zero degree. The 3dB beamwidth for a linear array with $K_s$ antennas is approximately $2\arcsin(\frac{1.2}{K_s})$ in radian. We generate the basic beams with $K_s=16$ and $12$ for the fixed and scanning subbeams, respectively. The cases of overlapping fixed and scanning subbeams are studied. The power distribution factor $\rho$ is set as 0.5. \cb{For communication, we consider a narrowband Rician channel, where the mean power ratio between the dominating LOS signal and NLOS signals is 10 dB. Consider the narrowband assumption, the difference in signal propagation delay is set to be negligible in communications, i.e., $\tau_l=0$. In the simulations, all the NLOS multipath components are randomly and uniformly distributed within an angular range of 14 degrees centered in the LOS direction. The total number of paths $L$ is 8, unless specified otherwise.} All the results of the received signal power for communications are normalized to the power value when the whole transmitter array generates a single beam pointing to the dominating AoD. To obtain the MSE of the BF waveform, the squared Euclidean norm of the difference between the generated BF radiation pattern and the desired one is averaged over randomly generated channel matrices.

For the methods proposed developed in Section \ref{sec-OptPhi}, $\wf_{t,c}$ points to the dominating AoD, and $\wf_{t,s}$ is generated by multiplying a phase-shifting sequence to the basic scanning subbeam to change the pointing directions, as described in \cite{zhang2019multibeam}. In the cases where the integral of the total scanning power needs to be calculated by \eqref{eq-integral}, we let $N_I=16$. It is observed that when $N_I\geq12$, each element in $\bm{\mathcal{A}}$ can achieve smaller errors than $10^{-3}$, compared to the value after convergence. The BF radiation pattern achieved by these values are nearly identical. 

For the methods developed in Section \ref{sec-SDR}, the MATLAB CVX toolbox is used and the $\text{SDPT}_3$ solver with default precision is employed. $\epsilon=\sqrt{\epsilon_0}$, where $\epsilon_0=2.22\times10^{-16}$ is the machine precision \cite{grant2008cvx}. The number of iterations $L_\text{max}$ is set to $5$. The values of the thresholds $\varepsilon_w$, $\varepsilon_{s_i}$, and $\varepsilon_p$ are set to be the product between a scalar in $[0.5, 1]$ and the MSE of the BF waveform achieved by Methods 2 in \cite{zhang2019multibeam}. 

With the above simulation settings, the computational complexity for the proposed methods in Section \ref{sec-OptPhi} is $O(16^2)$, and the worst-case complexity of the methods developed in Section \ref{sec-SDR} is $O(20\sqrt{2}\times16^{5}\times(\log2.22+8))\approx O(2.475\times10^8)$. 

\subsection{Results}

\begin{figure}[t]
	\centering
	\includegraphics[width=1\columnwidth]{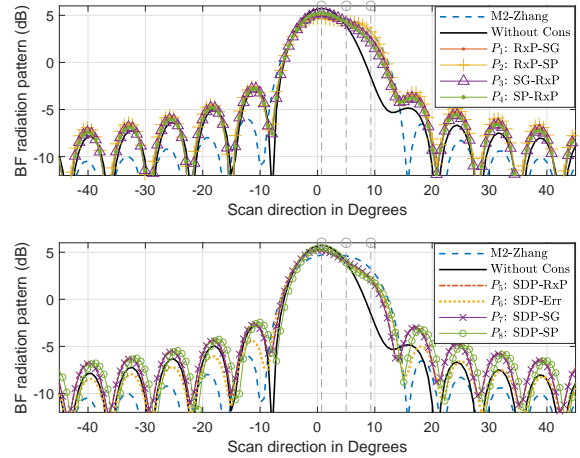}
	\caption{BF waveform (radiation pattern) when the scanning subbeam points at $5.01^\circ$. For ``MaxRxP-SG", ``MaxRxP-SP",``MaxSG-RxP", and ``MaxSP-RxP", $C_s=0.9$, $C_{sp}=0.9$, and $C_p=0.725$, respectively. For the methods constraining the power of the scanning subbeam, the integral range $(\theta_{s2}-\theta{s1})$ is $8.59^\circ$ (3dB beamwidth).}
	\label{fig-CorrectScan}
\end{figure}

Fig. \ref{fig-CorrectScan} shows the effectiveness of the proposed approaches in reducing the mismatches of the waveform. We can see that there can be a reduction of more than 4.5 dB in the gain in the desired scanning directions when only the received signal power is optimized, as compared to ``M2-Zhang". With multiple optimization objectives and constraints considered, the approaches proposed in this paper can achieve the BF waveforms much closer to the one using Method 2 in \cite{zhang2019multibeam}. \cb{Compared with ``M2-Zhang" which only optimizes the BF vector according to the desired BF waveform, the sidelobes of the BF waveform generated by the proposed methods are observed to slightly improve. This can disperse the power transmitted from the mainlobe and increase the signal power in undesired directions. Nevertheless, the proposed methods can balance between the performance of communication and sensing. The sidelobes can also be suppressed by imposing constraints on the desired BF waveform in these directions.}

\begin{figure}
	\centering \subfigure[For the methods constraining the gain at the dominating scanning direction, averaged normalized received signal power and MSE of the sensing BF waveform with varying $C_s$.]{ \label{fig-PowErr_Cs}	
		\includegraphics[width=0.9\columnwidth]{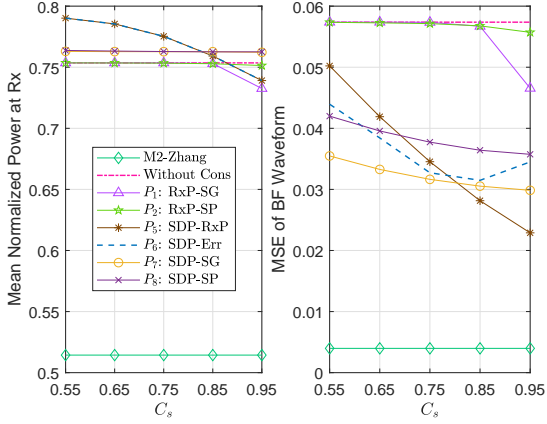}} 
	    \subfigure[For the methods using the constraint of the received signal power, averaged normalized received signal power and MSE of the sensing BF waveform with varying $C_p$.]{ \label{fig-PowErr_Cg} 
		\includegraphics[width=0.9\columnwidth]{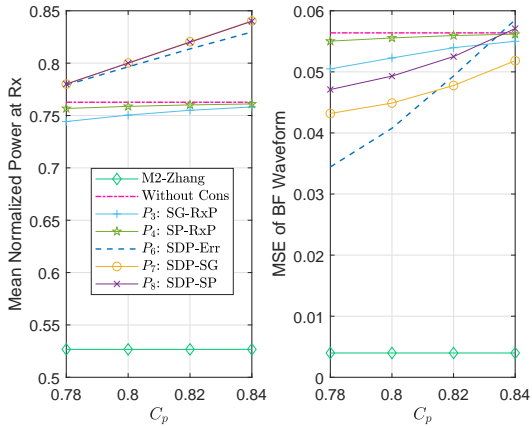}} 
	    \caption{For constrained multibeam generating methods, BF performance with varying bounds for the constraints. The scanning beam points to $-6.45^\circ$. } \label{fig-VaryBound} 
\end{figure}

Figs. \ref{fig-PowErr_Cs} and \ref{fig-PowErr_Cg} present how the values of the constraint thresholds influence the BF performance. The figures show that an increased threshold of the received power for communication generally results in a larger MSE of the sensing BF waveform, and a decreased threshold of the received power results in a smaller MSE. We also observe that, compared with the subbeam-combination methods, the global BF optimization generally achieves a better overall performance. \cb{For example, when $C_s\geq 0.85$ or $C_p < 0.84$, the global BF optimization methods achieve a higher received signal power and smaller MSEs of the BF waveform than the \textit{subbeam-combination} methods, for any given value of $C_s$ or $C_p$.}

\begin{figure}[t]
	\centering
	\includegraphics[width=1\columnwidth]{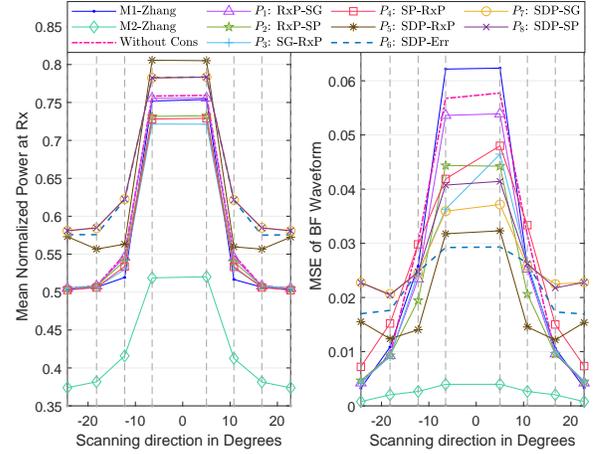}
	\caption{Normalized received signal power for communications and MSE of the scanning BF waveform for different BF methods when the scanning subbeam points to various directions. The scanning subbeams point to $-24.36^\circ$, $-18.21^\circ$, $-12.27^\circ$, $-6.45^\circ$, $5.02^\circ$, $10.81^\circ$, $16.71^\circ$, $22.80^\circ$, respectively. The values of $C_s$, $C_{sp}$, $C_p$ and $(\theta_{s2}-\theta{s1})$ are the same with those in Fig. \ref{fig-CorrectScan}. }
	\label{fig-PowErr}
\end{figure}

In Fig. \ref{fig-PowErr}, we show the normalized received signal power and the MSE of the sensing BF waveform in several different scanning directions. From the two subfigures, we can see that the global optimization methods achieve $5\%-10\%$ higher received signal powers than the subbeam-combination methods, with a reduced MSE of the BF waveform. For the subbeam-combination methods, the constrained methods lead to a slightly decreased received power, but better BF waveform, as compared to the unconstrained counterparts. 
When the fixed and scanning subbeams overlap substantially, the global optimization methods achieve a significantly lower waveform MSE (by up to approximately $50\%$, as compared to the unconstrained case), while maintaining a high received signal power. As typically expected, the waveform MSEs are larger when the constraints are imposed to the received signal power (i.e., solutions to $\text{P}_3$, $\text{P}_4$, $\text{P}_7$, and $\text{P}_8$).

\begin{figure}[t]
	\centering
	\includegraphics[width=0.9\columnwidth]{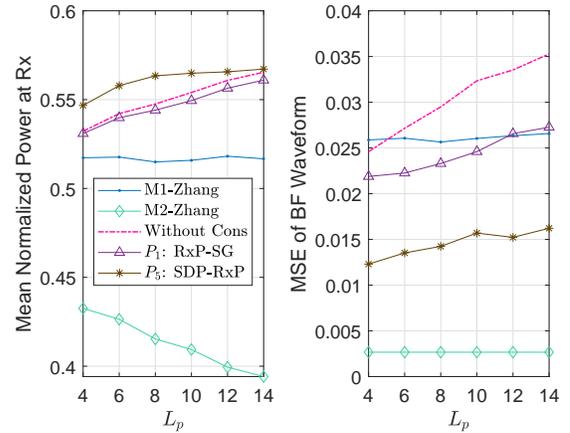}
	\caption{Normalized received signal power and MSE of BF waveform with varying number of paths $L$ when the scanning beam points to $-18.21^\circ$. The other settings are the same with those for Fig. \ref{fig-PowErr}. For ``$\text{P}_1$: RxP-SG", $C_s=0.9$. }
	\label{fig-varyL}
\end{figure}

In Fig. \ref{fig-varyL}, we show how the BF performance is affected by the number of NLOS paths. When the scanning subbeam points to $-12.18^\circ$, which means some paths may not be within the 3dB beamwidth of both fixed and scanning subbeams, the waveform MSE increases with the growth of $L$. For ``$\text{P}_5$: SDP-RxP", the MSE of the scanning BF waveform is even smaller than the other two subbeam-combination methods, although the received signal power of ``$\text{P}_5$: SDP-RxP" is higher. Similar results can be observed for the other methods and in other directions, which are not shown for the clarity of this figure. The figure also shows that the global optimization methods can balance (and better control) the different aspects of the BF performance.

\section{Conclusions}\label{sec-conc}

We studied a range of multibeam optimization methods for JCAS systems using analog arrays, considering the requirements of both communication and sensing. We first proposed new constrained optimization methods which provide closed-form optimal solutions to the phase coefficient for combining fixed and scanning BF vectors. We also proposed new global optimization methods that directly generate the single BF vectors. We presented the process of converting the original NP-hard problems to QCQP, which can be solved efficiently by using SDP techniques. The global optimization methods provide effective benchmarks for evaluating the performance tradeoff of other methods. Simulation results show that the proposed optimization methods can achieve a good balance between communication and sensing performances. 

\cb{Our work can be potentially extended to more complicated signal and channel models with frequency selectivity and beam squint effect \cite{wang2019beam}. For example, by referring to the method in \cite{wang2019beam}, one can formulate a cost function capturing all subcarriers, which considers both beam squint and frequency selectivity effects; and then optimize a single analog beamforming vector to minimize the cost function. The work in this paper can also be extended to hybrid arrays with multiple analog subarrays and RF chains, offering the capability of fine-tuning individual analog subarrays to improve digital beam synthesis of an entire hybrid array. }

\begin{appendices}
	
\cb{\section{Monotonicity analysis of $f(\varphi)$}\label{apdx-monotonicity}
	Referring to the derivation process in \cite{luo2019optimization}, the monotonicity of $f(\varphi)$ can be obtained by analyzing the sign of its first-order derivative 
	\begin{align}\label{eq-fvar}
	\begin{split}
	&f'(\varphi)=\frac{g_1'(\varphi)g_2(\varphi)-g_2'(\varphi)g_1(\varphi)}{g_2(\varphi)^2},\\
	\end{split}
	\end{align}
	where
	\begin{align*}
	&g_1(\varphi)\triangleq\rho\|\Ht\wcf\|^2+(1-\rho)\|\Ht\wsf\|^2\\
	&\qquad\quad\ +Pe^{j\varphi}\wcf^H\Ht^H\Ht\wsf+Pe^{-j\varphi}\wsf^H\Ht^H\Ht\wcf,\notag\\
	&g_2(\varphi)\triangleq\rho\|\wcf\|^2+(1-\rho)\|\wsf\|^2+Pe^{j\varphi}\wcf^H\wsf\\
	&\qquad \quad \ +Pe^{-j\varphi}\wsf^H\wcf.
	\end{align*}
	Obviously, $g_2^2(\varphi)>0$, which implies that we can determine the sign of $f'(\varphi)$ by analyzing the sign of the numerator in \eqref{eq-fvar}. Let $h(\varphi)$ be the numerator, and let $\wcf^H\Ht^H\Ht\wsf=a_1e^{j\alpha_1}$ and $\wcf^H\wsf=a_2e^{j\alpha_2}$, where $a_1\geq 0$ and $a_2\geq 0$. We have
	\begin{align*}
	h(\varphi)&=-2Pa_1\sin(\varphi+\alpha_1)-4P^2a_1a_2\sin(\alpha_1-\alpha_2)+\notag\\
	&\quad\ 2Pa_2[\rho\|\Ht\wcf\|^2+(1-\rho)\|\Ht\wsf\|^2]\sin(\varphi+\alpha_2)\notag\\
	&=X_1\sin(\varphi)+X_2\cos(\varphi)+L,
	\end{align*}
	where 
	\begin{align*}
	X_1\triangleq&2P|a_1|\cos\alpha_1+2P|a_2|[\rho\|\Ht\wcf\|^2\\
	&+(1-\rho)\|\Ht\wsf\|^2]\cos\alpha_2,\notag\\
	X_2\triangleq&-2P|a_1|\sin\alpha_1+2P|a_2|[\rho\|\Ht\wcf\|^2\\
	&+(1-\rho)\|\Ht\wsf\|^2]\sin\alpha_2,\notag\\
	L\triangleq&-4P^2|a_1||a_2|\sin(\alpha_1-\alpha_2).
	\end{align*}
	By considering the sign of $X_1$, $h(\varphi)$ can be written as
	\begin{align*}
	h(\varphi)=\left\{\begin{array}{cc}
	\sqrt{X_1^2+X_2^2}\sin(\varphi+\zeta)+L, &\text{if}\ X_1\geq 0\\
	-\sqrt{X_1^2+X_2^2}\sin(\varphi+\zeta)+L, & \text{if}\ X_1<0,
	\end{array}\right.
	\end{align*}
	where $\zeta=\arctan(X_2/X_1)$. 
	
	Since $h(\varphi)$ is a cyclic function and a cycle lasts $2\pi$, we study the monotonicity of $f(\varphi)$ in one cycle. Details that can be found in \cite{luo2019optimization} are omitted due to page limit. When $X_1>0$, the monotonic intervals is summarized in Table \ref{tb-monotonyvarphi}, where $\mu_0=\arcsin{(\dfrac{L}{\sqrt{X_1^2+X_2^2}})}$. When $X_1<0$, the monotonicity of $f(\varphi)$ is opposite to what it is when $X_1>0$.    
	
	\begin{table*}[!t]
		\renewcommand{\arraystretch}{2.0}
		\caption{For $X_1>0$, the monotonicity and maximum of $f(\varphi)$ in one period. }
		\label{tb-monotonyvarphi}
		\centering
		\resizebox{2.0\columnwidth}{!}{%
			\subtable[$L>0$]{
				\begin{tabular}{|c|c|c|c|c|c|c|}
					\hline  \textbf{Range of $\varphi$}  &$(-2\pi-\mu_0-\zeta,-\pi+\mu_0-\zeta)$ &$-\pi+\mu_0-\zeta$ &$(-\pi+\mu_0-\zeta,-\mu_0-\zeta)$ &$-\mu_0-\zeta$ & $(-\mu_0-\zeta,\pi+\mu_0-\zeta)$ & $\pi+\mu_0-\zeta$\\ 
					\hline\textbf{ Sign of $f'(\varphi)$ }& $>0$ & $0$ & $<0$ &$0$ &$>0$ & $0$ \\ 
					\hline \textbf{$f(\varphi)$} & monotonically increasing & maximum & monotonically decreasing & minimum & monotonically increasing& maximum\\	\hline
				\end{tabular}}	}
				\resizebox{2.0\columnwidth}{!}{%
					\subtable[$L<0$]{
						\begin{tabular}{|c|c|c|c|c|c|c|}
							\hline  \textbf{Range of $\varphi$} &$(-\pi+\mu_0-\zeta,-\mu_0-\zeta)$ &$-\mu_0-\zeta$ &$(-\mu_0-\zeta,\pi+\mu_0-\zeta)$ & $\pi+\mu_0-\zeta$&$(\pi+\mu_0-\zeta,2\pi-\mu_0-\zeta)$ & $\cdots$\\ 
							\hline\textbf{ Sign of $f'(\varphi)$} & $<0$ & $0$ & $>0$ & $0$ & $<0$ & $\cdots$\\ 
							\hline $f(\varphi)$ & monotonically decreasing & minimum & monotonically increasing & maximum & monotonically decreasing& $\cdots$\\ 
							\hline 
						\end{tabular} }	}
					\end{table*}
				}

\section{The Range of $\varphi$ under Constraints \eqref{eq-stScanPow}} \label{apdx-rangeInt}

We rewrite \eqref{eq-stScanPow} as
\begin{align}\label{eq-ineqPow}
\frac{h_{p_1}(\varphi)}{h_{p_2}(\varphi)}\geq &C_{sp}\wf_2^H\bm{\mathcal{A}}\wf_2 \text{ (or }C_{sp_2}\wf_s^H\bm{\mathcal{A}}\wf_s \text{)},
\end{align}
where
\begin{align}
h_{p_1}(\varphi)=&\rho\wcf^H\bm{\mathcal{A}}\wcf+(1-\rho)\wsf^H\bm{\mathcal{A}}\wsf\notag\\
&+2P\mathfrak{Re}\{e^{j\varphi}\wcf^H\bm{\mathcal{A}}\wsf\},\\
h_{p_2}(\varphi)=&\rho\|\wcf\|^2+(1-\rho)\|\wsf\|^2+2P\mathfrak{Re}\{e^{j\varphi}\wcf^H\wsf\}\notag\\
=&1+2P\mathfrak{Re}\{e^{j\varphi}\wcf^H\wsf\}.
\end{align}
Let $\wcf^H\bm{\mathcal{A}}\wsf=b_{p}e^{j\beta_{p}}$, $B_{p_1}\triangleq[\rho\wcf^H\bm{\mathcal{A}}\wcf+(1-\rho)\wsf^H\bm{\mathcal{A}}\wsf]/2P$, and $B_{p_2}\triangleq C_{sp}\wf_2^H\bm{\mathcal{A}}\wf_2/2P$, and \eqref{eq-ineqPow} can be converted to 
\begin{align}
\left\{
\begin{array}{l}
C_{p_1}\sin\varphi+C_{p_2}\cos\varphi\geq {B_{p_2}-B_{p_1}}, \\
C_{p_1}\triangleq 2Pb_1B_{p_2}\sin\beta_1-b_p\sin\beta_p,\\
C_{p_2}\triangleq b_p\cos\beta_p-2Pb_1B_{p_2}\cos\beta_1.
\end{array}
\right.
\end{align}
Consider the following three cases:
\begin{enumerate}
	\item If $|{B_{p_2}-B_{p_1}}|\leq{\sqrt{C_{p_1}^2+C_{p_2}^2}}$, we can obtain
\begin{align*}
\begin{split}
\varphi&\in\Bbbk_p= \left[\varphi_{p1},\varphi_{p2}\right]\\
&=\left\{\begin{array}{cl}
\left[\mu_p-\sigma_p,-\mu_p+\pi-\sigma_p\right], \text{if}\ C_{p_1}\geq 0, \\
\left[\mu_p+\pi-\sigma_p,-\mu_p+2\pi-\sigma_p\right], \text{if}\ C_{p_1}<0,
\end{array}\right.
\end{split}
\end{align*}
where $\mu_p \triangleq \arcsin(\frac{B_{p_2}-B_{p_1}}{\sqrt{C_{p_1}^2+C_{p_2}^2}})+2k\pi, \ k=\pm 1,\pm 2,\cdots$ and $\sigma_p\triangleq \arctan(\frac{C_{p_2}}{C_{p_1}})$. 
\item If ${B_{p_2}-B_{p_1}}\leq{-\sqrt{C_{p_1}^2+C_{p_2}^2}}$, we have $\Bbbk_p=\mathbf{R}$. 
\item If ${B_{p_2}-B_{p_1}}>{\sqrt{C_{p_1}^2+C_{p_2}^2}}$, we have $\Bbbk_p\in\varnothing$. 
\end{enumerate}

\section{The Range of $\varphi$ under Constraints \eqref{eq-stOutPow}}\label{apdx-phig}

Start with expanding the left-hand side of \eqref{eq-stOutPow}. The expansion of its denominator is the same as it is with \eqref{eq-ineq}, and for the numerator we have
\begin{align}
&\wtf^H\Ht^H\Ht\wtf\\ \notag
&=\rho\|\Ht\wcf\|^2+(1-\rho)\|\Ht\wsf\|^2 \notag\\
&\ \  +Pe^{j\varphi}\wcf^H\Ht^H\Ht\wsf+Pe^{-j\varphi}\wsf^H\Ht^H\Ht\wcf.
\end{align}
Let $\wcf^H\Ht^H\Ht\wsf=b_ge^{j\beta_g}$, $B_{g_1}\triangleq[\rho\|\Ht\wcf\|^2+(1-\rho)\|\Ht\wsf\|^2]/2P$, and $B_{g_2}\triangleq C_pP_c/2P$. Then \eqref{eq-stOutPow} can be converted to 
\begin{align}
\left\{
\begin{array}{l}
C_{g_1}\sin\varphi+C_{g_2}\cos\varphi\geq {B_{g_2}-B_{g_1}}, \\
C_{g_1}\triangleq 2Pb_1B_{g_2}\sin\beta_1-b_g\sin\beta_g,\\
C_{g_2}\triangleq b_g\cos\beta_g-2Pb_1B_{g_2}\cos\beta_1.
\end{array}
\right.
\end{align}
Considering the three cases similar to those in Section \ref{sec-ScanGain} and Appendix \ref{apdx-rangeInt}, we can specify the range of $\varphi$ as  
\begin{align}\label{eq-kg}
\begin{split}
\Bbbk_g=\left\{\begin{array}{cc}\left[\varphi_{g1},\varphi_{g2}\right], &\text{ if } |{B_{g_2}-B_{g_1}}|\leq{\sqrt{C_{g_1}^2+C_{g_2}^2}},\\
\bm{R}, &\text{ if }{B_{g_2}-B_{g_1}}\leq{-\sqrt{C_{g_1}^2+C_{g_2}^2}},\\
\varnothing, &\text{ if }{B_{g_2}-B_{g_1}}>{\sqrt{C_{g_1}^2+C_{g_2}^2}}.\\
\end{array}\right.\\
\end{split}
\end{align}
where
\begin{align*}
\begin{split}
&\left[\varphi_{g1},\varphi_{g2}\right]\\
&=\left\{\begin{array}{cl}
\left[\mu_g-\sigma_g,-\mu_g+\pi-\sigma_g\right], &\text{if}\ C_{g_1}\geq 0, \\
\left[\mu_g+\pi-\sigma_g,-\mu_g+2\pi-\sigma_g\right], &\text{if}\ C_{g_1}<0,
\end{array}\right. \\
\end{split}
\end{align*}
and 
\begin{align*}
&\mu_g\triangleq\arcsin(\frac{B_{g_2}-B_{g_1}}{\sqrt{C_{g_1}^2+C_{g_2}^2}})+2k\pi, \ k=\pm 1,\pm 2,\cdots\\
&\sigma_g\triangleq\arctan(\frac{C_{g_2}}{C_{g_1}}).
\end{align*}

\section{Equivalence of complex-valued and real-valued optimization problems}\label{apdx-Complex2Real}	
Separating the real part from the imaginary, $\At$, $\wtf$, and $\df_v$ can be written as
\begin{align}\label{eq-SepReIm}
\begin{split}
&\At=\mathfrak{Re}\{\At\}+j\mathfrak{Im}\{\At\}=\At_R+j\At_I,\\ 
&\wtf=\mathfrak{Re}\{\wtf\}+j\mathfrak{Im}\{\wtf\}=\wf_{tR}+j\wf_{tI},\\
&\df_v=\mathfrak{Re}\{\df_v\}+j\mathfrak{Im}\{\df_v\}=\df_{vR}+j\df_{vI}.
\end{split}
\end{align}	
Substituting \eqref{eq-SepReIm} into constraint \eqref{eq-WVFM}, we can obtain 
\begin{align}\label{eq-ReSepIm}
\begin{split}
&\|\Dt\At\wtf\|^2\\
&=(\wf_{tR}^T-j\wf_{tI}^T)(\At_R^T-j\At_I^T)\Dt^T\Dt(\At_R+j\At_I)(\wf_{tR}+j\wf_{tI})\\
&=(\wf_{tR}^T\At_R^T-\wf_{tI}^T\At_I^T)\Dt^T\Dt(\At_R\wf_{tR}-\At_I\wf_{tI})+\\
&\quad\ (\wf_{tI}^T\At_R^T+\wf_{tR}^T\At_I^T)\Dt^T\Dt(\At_R\wf_{tI}-\At_I\wf_{tR}),\\
&{\mathfrak{Re}^2\{\df_v^H\Dt^H\Dt\At\wtf\}}/{\|\df_v\|^2}\\
&=\mathfrak{Re}^2\{(\df_{vR}^T-j\df_{vI}^T)\Dt^T\Dt(\At_R+j\At_I)(\wf_{tR}+j\wf_{tI})\}\\
&=|(\df_{vR}^T\Dt^T\Dt\At_R+\df_{vI}^T\Dt^T\Dt\At_I)\wf_{tR}+\\
&\quad\ (\df_{vI}^T\Dt^T\Dt\At_R-\df_{vR}^T\Dt^T\Dt\At_I)\wf_{tI}|^2/{(\|\df_{vR}\|^2+\|\df_{vI}\|^2)}.
\end{split}
\end{align}
Using $\tilde{\At}$, $\tilde{\wf}_t$, $\tilde{\Dt}$, and $\tilde{\df}_v$ defined in \eqref{eq-real}, it is easy to verify that
\begin{align}
\begin{split}
&|\tilde{\Dt}^T\tilde{\At}\tilde{\wf}_t|^2=(\wf_{tR}^T\At_R^T-\wf_{tI}^T\At_I^T)\Dt^T\Dt(\At_R\wf_{tR}-\At_I\wf_{tI})+\\
&\qquad\qquad\qquad(\wf_{tI}^T\At_R^T+\wf_{tR}^T\At_I^T)\Dt^T\Dt(\At_R\wf_{tI}-\At_I\wf_{tR}),\\
&|\tilde{\df}_v^T\tilde{\Dt}^T\tilde{\Dt}\tilde{\At}\tilde{\wf}_t|^2/\|\tilde{\df}_v\|^2\\
&=|(\df_{vR}^T\Dt^T\Dt\At_R+\df_{vI}^T\Dt^T\Dt\At_I)\wf_{tR}+\\
&\quad\ (\df_{vI}^T\Dt^T\Dt\At_R-\df_{vR}^T\Dt^T\Dt\At_I)\wf_{tI}|^2/{(\|\df_{vR}\|^2+\|\df_{vI}\|^2)}.
\end{split}
\end{align}
After mathematical manipulation, $|\tilde{\Dt}^T\tilde{\At}\tilde{\wf}_t|^2-|\tilde{\df}_v^T\tilde{\Dt}^T\tilde{\Dt}\tilde{\At}\tilde{\wf}_t|^2/\|\tilde{\df}_v\|^2$ can be converted to the form of the waveform constraint \eqref{eq-realConsWVFM}. The equivalence between the objective function and the other constraints in \eqref{eq-formulation2} and \eqref{eq-realPro2} can be established in the same way.

\end{appendices}
\bibliographystyle{IEEEtran}
\bibliography{IEEEabrv,Yuyue1}

\end{document}